\documentclass[a4paper]{article}
\usepackage{Odyssey2022}
\usepackage{epsfig,amssymb,amsmath}
\usepackage{xpatch,xcolor}
\usepackage{arydshln}
\ninept
\makeatletter
\def\name#1{\gdef\@name{#1\\}}
\makeatother
\usepackage{footnote}
\makesavenoteenv{tabular}
\makesavenoteenv{table}
\usepackage[symbol]{footmisc}
\usepackage[titletoc]{appendix}
\usepackage{cite,url}
\usepackage{hyperref,xcolor}
\hypersetup{
    colorlinks=true,
    linkcolor=purple,
    filecolor=magenta,      
    urlcolor=purple,
}

\title{Automatic speaker verification spoofing and deepfake detection \\ using wav2vec 2.0 and data augmentation}

\name{
{\em Hemlata Tak$^1$, Massimiliano Todisco$^1$, Xin Wang$^2$, Jee-weon Jung$^3$}\\ {\em Junichi Yamagishi$^2$ and Nicholas Evans$^1$}
\thanks{The first author is supported by the VoicePersonae project funded by the French Agence Nationale de la Recherche (ANR) and the Japan Science and Technology Agency (JST).}}

\address{$^1$EURECOM, France, $^2$ National Institute of Informatics, Japan\\  $^3$Naver Corporation, South Korea \\
{\small \tt \{tak,todisco,evans\}@eurecom.fr, \{wangxin,jyamagis\}@nii.ac.jp, jeeweon.jung@navercorp.com} }

\begin{document}
\maketitle

\begin{abstract}

The performance of spoofing countermeasure systems depends fundamentally upon the use of sufficiently representative training data. With this usually being limited, current solutions typically lack generalisation to attacks encountered in the wild. Strategies to improve reliability in the face of uncontrolled, unpredictable attacks are hence needed. We report in this paper our efforts to use self-supervised learning in the form of a wav2vec 2.0 front-end with fine tuning. Despite initial base representations being learned using only bona fide data and no spoofed data, we obtain the lowest equal error rates reported in the literature for both the ASVspoof 2021 Logical Access and Deepfake databases. When combined with data augmentation, these results correspond to an improvement of almost 90\% relative to our baseline system.

\end{abstract}

\section{Introduction}

A persisting challenge in the design of spoofing countermeasures (CMs) for automatic speaker verification (ASV) is reliability in the face of diverse, unpredictable attacks~\cite{nautsch2021asvspoof}. ASV systems can be compromised by attacks belonging to a broad variety of different classes, e.g., converted voice, synthetic speech and replayed recordings.  Even within each attack class, there is considerable potential variation, e.g., different algorithms or recording and replay device characteristics.  An ideal spoofing detection solution should be robust to all such variation even if, in the wild, it is unpredictable.  The acquisition of training data that is representative of spoofing attacks with near-boundless variability is obviously impracticable.

The ASVspoof initiative and challenge series have collected large databases of spoofed and bona fide utterances that are suitable for the training of spoofing countermeasures.  To promote the development of generalisable countermeasures, namely detection solutions that cope well in the face of spoofing attacks not previously encountered, assessment is performed with experimental protocols and evaluation data that comprise spoofed utterances generated with a broad variety of different algorithms or techniques.  The differences between training, development and evaluation data can lead to substantial differences in detection performance.  For the most recent ASVspoof 2021 Logical Access (LA) evaluation~\cite{yamagishi2021_ASV_spoof}, the equal error rate (EER) of the best performing baseline solution increased from 0.55\% for the development set to  9.26\% for the evaluation set~\cite{yamagishi2021_ASV_spoof}.  Submission results show better performance~\cite{tomilov2021stc,chen21b_asvspoof,kang2021crim,caceres2021biometric,das2021known,chen21_asvspoof,wang21_ASV_spoof}, but the fundamental gap between performance for development and evaluation data remains, indicating a persisting lack of generalisation.

\begin{figure*}[!ht]
  \centering
  \includegraphics[trim={6.4cm 4.5cm 3.6cm 5cm},clip,width=1\linewidth]{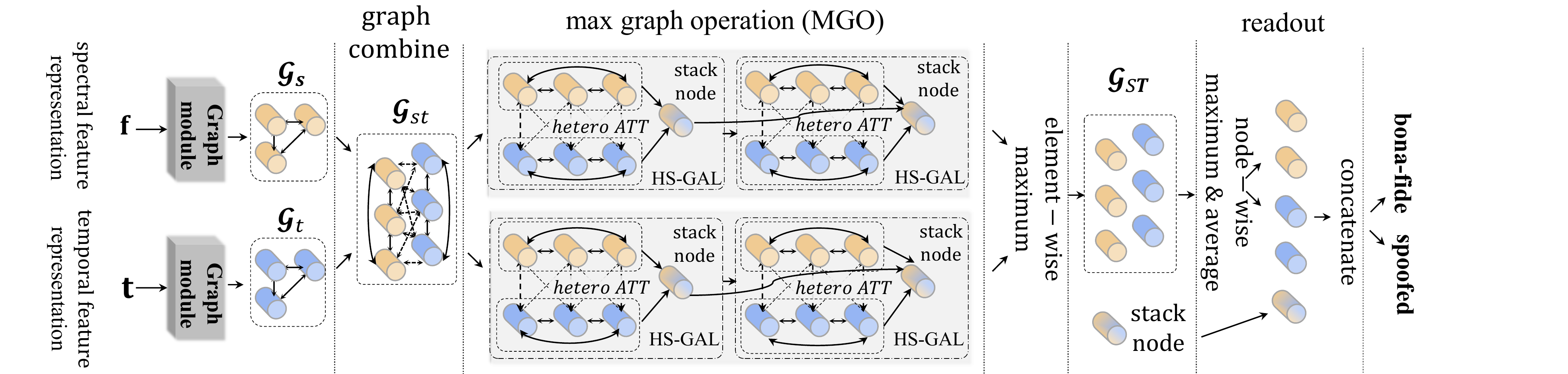}
 \caption{AASIST baseline framework reproduced from~\cite{jung2021aasist}. }
   \label{fig:baseline framework}
 \end{figure*}
 
 \begin{figure}[!ht]
  \centering
  \includegraphics[trim={5cm 1cm 11.0cm 0.0cm},clip,width=1.7\linewidth]{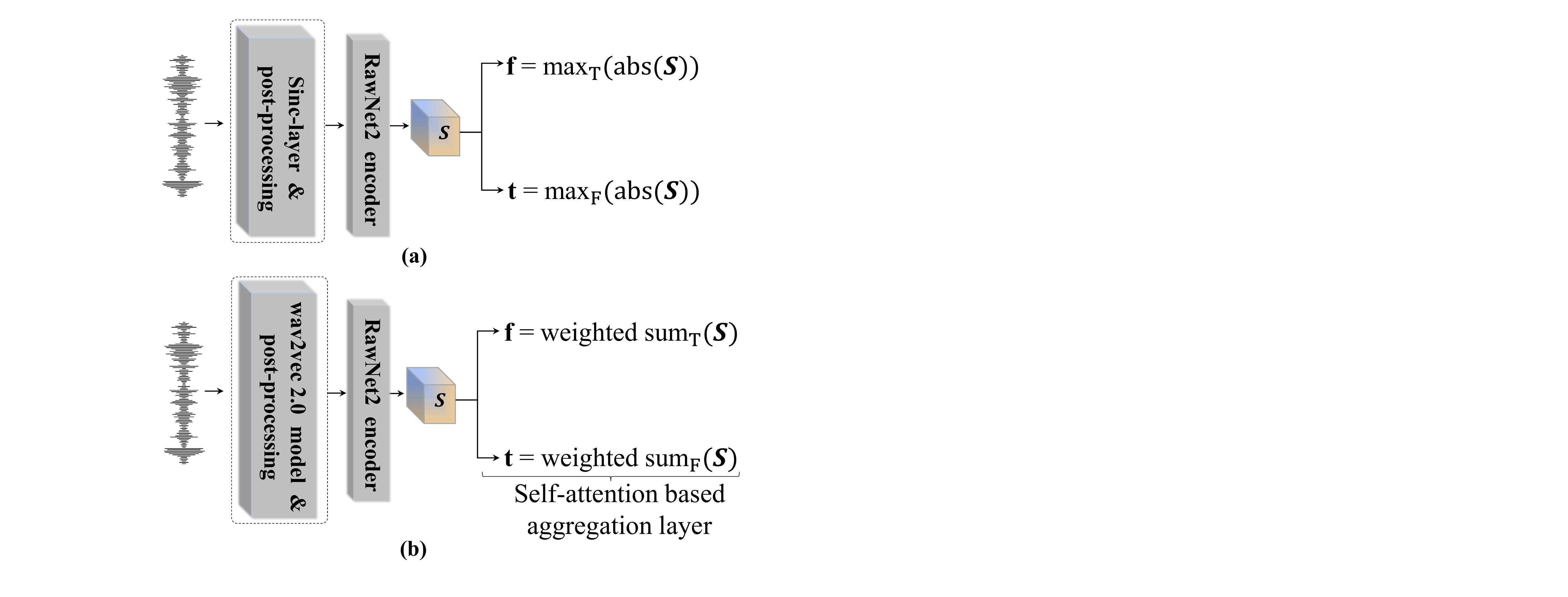}
 \caption{Front-end systems: (a) the baseline sinc-layer front-end; (b) the  wav2vec 2.0 front-end.}
   \label{fig:front-end framework}
 \end{figure}

Given that the training data used for ASVspoof challenges comprises spoofed utterances generated with a modest number of different attack algorithms (six in the case of the ASVspoof 2019 LA database), the lack of generalisation may be difficult to improve upon unless challenge rules are relaxed to allow training using external data. 
A relaxed training data policy would complicate comparisons between different systems and technology trained using different data -- the playing field would no longer be level -- though potential improvements to generalisation may make it worthwhile. 

The question then is what external training data to use and how to use it?  With the drive toward reproducible research, a number of speech synthesis and voice conversion algorithms are now openly available as open source.  Additional training data, generated with different attack algorithms, can hence be produced readily.  The number of algorithms remains limited, however, and can fundamentally never be fully representative of what can reasonably be expected in the wild.  We have hence explored a different approach.

Motivated by (i)~its proven application to the learning of general neural representations for a range of different tasks~\cite{van2018representation,baevski2020wav2vec,babu2021xls,yang2021superb,conneau2020unsupervised,kawakami2020learning,cai2021iterative,tjandra2021improved,evain2021lebenchmark}, (ii)~evidence that fine-tuning with modest quantities of labelled data leads to state-of-the-art results, (iii)~encouraging, previously reported results for anti-spoofing~\cite{xie2021siamese,wang2021investigating} and (iv)~the appeal of one-class classification approaches~\cite{alegre2013one}, we have explored the use of self-supervised learning to improve generalisation.  Our hypothesis is that better representations trained on diverse speech data, even those learned for other tasks and initially using only bona fide data (hence one-class), may help to reduce over-fitting and hence improve reliability and domain-robustness, particularly in the face of previously unseen spoofing attacks.  Additionally, we hope that better trained representations are complementary to data augmentations techniques which are already known to improve generalisation~\cite{chen2020generalization,das2021data,das2021known,zhang2021empirical,chen2021ur,tak2021rawboost}.

The principal contributions of this work are: (i)~improved generalisation and domain robustness using a pre-trained, self-supervised speech model with fine-tuning; (ii)~additional improvements using data augmentation showing complementary benefits to self-supervised learning; %(iii)~improvements to domain robustness; 
(iii)~a new self-attention based aggregation layer which brings complementary improvements.

\begin{figure*}[!t]
  \centering
  \includegraphics[trim={3.0cm 2.4cm 5.5cm 3.4cm},clip,width=0.775\linewidth]{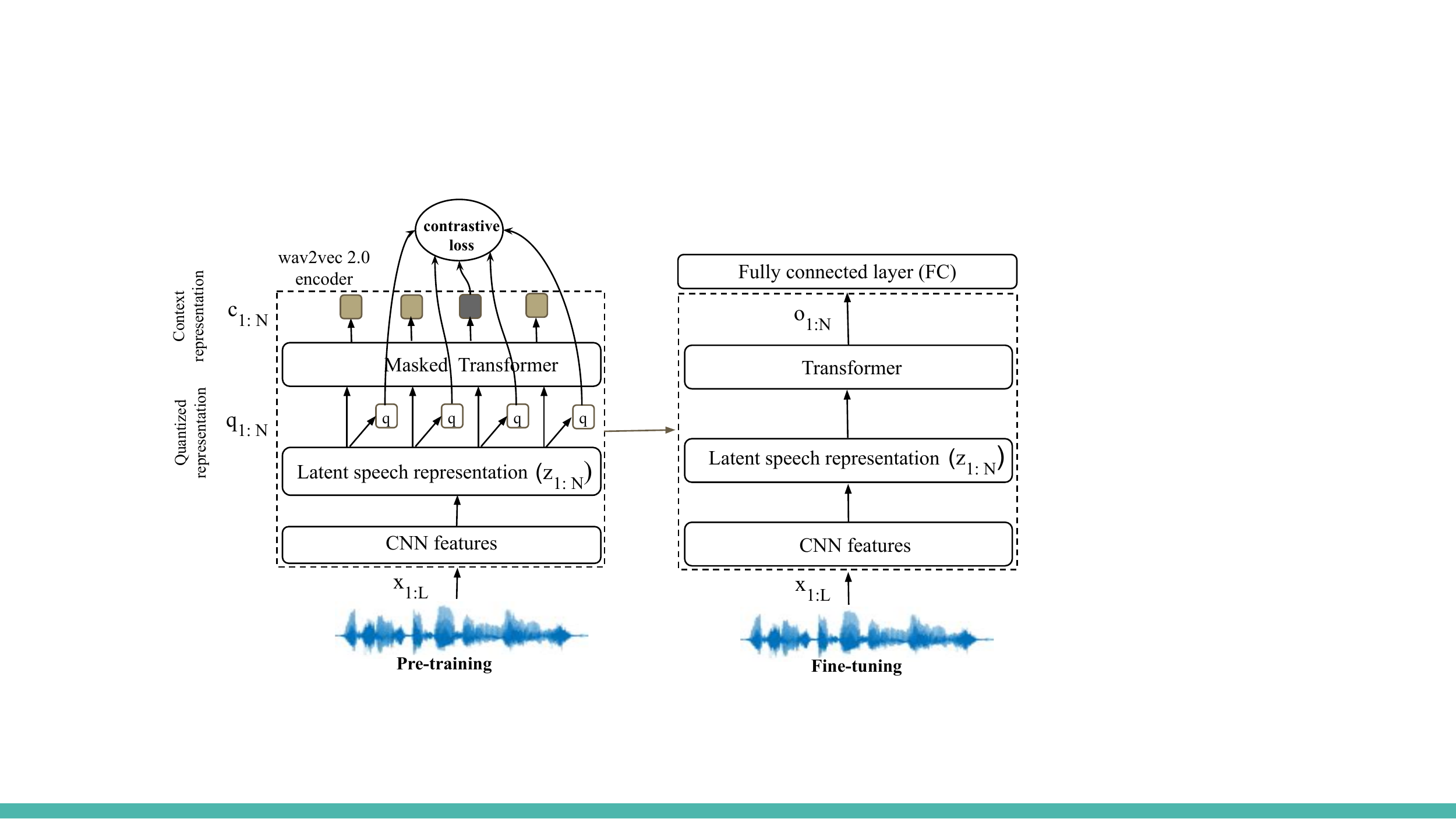}
 \caption{An overview of the pre-training and fine-tuning of the wav2vec 2.0 model, adapted from~\cite{babu2021xls}. }
  \label{fig:wav2vec framework}
\end{figure*}

\section{Related work}

Self-supervised learning (SSL) has attracted considerable attention in recent times.  Existing works show that pre-trained models derived using SSL generalise well across a multitude of different tasks when a relatively simple classifier is learned from the resulting representations using only a modest quantity of labelled data~\cite{kenton2019bert}.  A growing number of self-supervised speech models have been proposed.  Examples include contrastive predictive coding (CPC)~\cite{oord2018representation,kawakami2020learning}, auto-regressive predictive coding~\cite{chung2021similarity}, wav2vec~\cite{schneider2019wav2vec}, HuBERT~\cite{hsu2021hubert,hsu2021hubert1}, wav2vec 2.0~\cite{baevski2020wav2vec,xu2021self} and Wavlm~\cite{chen2021wavlm}, with all showing promising results for a variety of different speech processing tasks.

Two particularly popular approaches, HuBERT and wav2vec 2.0, have been applied to automatic speech recognition~\cite{baevski2020wav2vec,babu2021xls}, mispronunciation detection~\cite{xu2021explore,peng2021study}, speaker recognition~\cite{vaessen2021fine,fan2020exploring} and emotion recognition~\cite{pepino2021emotion}. The same techniques have been explored in the context of spoofing detection~\cite{xie2021siamese,wang2021investigating}.
Xie et al.~\cite{xie2021siamese} showed the benefit of using SSL with a Siamese network for spoofing detection.  With learned representations coming from their combination, and without comparative assessments using representations derived via alternative means, the specific benefits of SSL are  difficult to judge.

Wang et al.~\cite{wang2021investigating} compared different SSL based front-ends and back-end architectures and showed the importance of fine-tuning SSL models for spoofing detection. By replacing a linear frequency cepstral coefficient front-end with a wav2vec 2.0 front-end and by fine-tuning, they achieved relative reductions in the equal error rate (EER) of 68\% and 79\% for the ASVspoof 2021 LA and DF databases respectively.  Nonetheless, the EERs remain at 5\% and the additional or complementary benefit of data augmentation, which is known to be beneficial in both cases~\cite{chen21b_asvspoof,tomilov2021stc,das2021known}, was not explored.  Results showed that the wav2vec 2.0~\cite{baevski2020wav2vec} front-end gives better generalised spoofing detection performance than a HuBERT model. 

We have hence explored the wav2vec 2.0 XLS-R  (0.3B) model\footnote{\url{https://github.com/pytorch/fairseq/tree/main/examples/wav2vec/xlsr}}~\cite{babu2021xls} in our work. wav2vec 2.0 XLS-R  is a large-scale cross-lingually pre-trained model trained on diverse corpora including VoxPopuli data (VP-400K)~\cite{wang2021voxpopuli}, the multilingual Librispeech corpus (MLS)~\cite{pratap2020mls}, CommonVoice (CV)~\cite{ardila2020common}, VoxLingua107 (VL)~\cite{valk2021voxlingua107}, and BABEL (BBL)~\cite{gales2014speech} datasets.  Together, they include speech data in 128 different languages from many different regions of the world. We explored the wav2vec 2.0 front-end with an integrated spectro-temporal graph attention network (AASIST) as a back-end~\cite{jung2021aasist}.  The latter is described in the next section.  Its coupling with a wav2vec 2.0 front-end is described in Section~\ref{sec:SSL_front_end}.

\section{AASIST baseline system}
\label{sec:baseline}

The baseline system is an end-to-end, integrated spectro-temporal graph attention network named AASIST~\cite{jung2021aasist}, illustrated in Fig.~\ref{fig:baseline framework}. It extracts representations directly from raw waveform inputs. As illustrated in Fig.~\ref{fig:front-end framework}-(a),
AASIST uses a sinc convolutional layer based front-end~\cite{ravanelli2018speaker}. It is initialised with 70 mel-scaled filters, each with a kernel size of 129~\cite{tak2021end}. Through the addition of a channel dimension, the sinc layer output is fed to a post-processing layer and transformed to a spectro-temporal representation. These are fed to a RawNet2-based residual encoder, to learn a higher-level feature map $\textbf{S}\in\mathbb{R}^{C \times F\times T}$ where $C$, $F$ and $T$ refer to the number of channels, spectral bins and time frames respectively. 

Separate spectral and temporal representations are learned from $\textbf{S}$ using a max-pooling operation which is applied to the absolute values across either temporal or spectral dimensions in order to construct either a spectral input graph ($\mathcal{G}_{s} \in\mathbb{R}^{N_\textbf{s} \times d_{s}} $) or a temporal input graph ($\mathcal{G}_{t} \in\mathbb{R}^{N_\textbf{t} \times d_{t}}$).  $N_s$ and $N_t$ are the set of graph nodes in spectral and temporal graphs respectively whereas $d$ is the feature dimensionality of each node.
Spectral and temporal graphs $\mathcal{G}_s$ and $\mathcal{G}_t$ are modelled using a pair of parallel graph modules (grey boxes in Fig.~\ref{fig:baseline framework}), each comprising a graph attention network (GAT)~\cite{velivckovic2017graph} and a graph-pooling layer~\cite{gao2019graph}: 

\begin{equation}
\mathcal{G}_t=\textrm{graph}\_\textrm{module}({\textrm{max}}_{F}(\textrm{abs}(\textbf{S}))) 
\end{equation}
\begin{equation}
\mathcal{G}_s =\textrm{graph}\_\textrm{module}({\textrm{max}}_{T}(\textrm{abs}(\textbf{S})))
\end{equation}

A heterogeneous spectro-temporal graph ($\mathcal{G}_{st}$) is then formed 
by combining temporal ($\mathcal{G}_{t}$) and spectral ($\mathcal{G}_{s}$) graphs using a heterogeneous stacking graph attention layer (HS-GAL).  Graph combination enables the concurrent modelling of heterogeneous graph representations with different node dimensions.
An HS-GAL contains an attention mechanism modified in order to accommodate graph heterogeneity~\cite{wang2019heterogeneous} and an additional stack node~\cite{kenton2019bert}. The latter acts to capture the relationships between artefacts spanning temporal and spectral domains. First, $\mathcal{G}_{t}$ and $\mathcal{G}_{s}$
are projected using an affine-transformation to another latent space 
with common dimensionality $d_{st}$ before being fed into the HS-GAL which then constructs a combined heterogeneous graph {$\mathcal{G}_{st}$}.\\

HS-GALs are applied with a max graph operation (MGO) where two branches, each consisting of two HS-GALs, learn to detect different spoofing artefacts in parallel. Each HS-GAL is followed by a graph pooling layer and an element-wise maximum operation is applied to the branch outputs to produce another heterogeneous graph {$\mathcal{G}_{ST}$}.  HS-GALs in each branch share a common stack node.  The stack node of each preceding HS-GAL is fed to the following HS-GAL so that information in both temporal and spectral graphs is preserved. 

The readout scheme (penultimate block in Fig.~\ref{fig:baseline framework}) uses node-wise maximum and average operations. The output of the readout layer is formed from the concatenation of five nodes.  The first four nodes are derived by applying a maximum and average to spectral nodes (orange) and temporal nodes (blue) in $\mathcal{G}_{ST}$. The fifth is the copied stack node. A two-class prediction output (bona fide and spoofed) is finally generated using a hidden fully connected layer.

\section{Self-supervised front-end}
\label{sec:SSL_front_end}
In this section we describe the replacement of the sinc-layer front-end 
with a wav2vec 2.0 model as illustrated in 
Fig.~\ref{fig:front-end framework}-(b). We describe both pre-training and fine-tuning to support downstream spoofing detection, both illustrated in Fig.~\ref{fig:wav2vec framework}.

\subsection{Wav2vec 2.0 model}

The wav2vec 2.0 pre-trained model is used to extract a sequence of feature representations $o_{1:N}$ from the raw input waveform $x_{1:L}$, where $L$ is the number of samples. As shown in Fig.~\ref{fig:wav2vec framework}, the wav2vec 2.0 model consists of a convolutional neural network (CNN) and a transformer~\cite{vaswani2017attention,kenton2019bert} network. The former converts the input $x_{1:L}$ to a hidden feature sequence $z_{1:N}$ whereas the latter transforms $z_{1:N}$ to output sequence $o_{1:N}$. The ratio between $L$ and $N$ is dictated by the CNN stride of 20~ms (the default setting).

\subsection{Pre-training}

An illustration of the pre-training procedure following~\cite{baevski2020wav2vec} is illustrated to the left in Fig.~\ref{fig:wav2vec framework}.
Latent representations $z_{1:N}$ are quantised to  representations $q_{1:N}$.  
Some portion of the latent representation $z_{1:N}$ is then masked and fed to the transformer which builds new context representations $c_{1:N}$.
A contrastive loss for each masked time step $n$ is then computed to measure how well the target $q_n$ can be identified from among a set of distractors (i.e., $q_{n'}$ sampled from the other masked time steps where $n'\neq n$) given the corresponding context vector $c_n$. All work reported in this paper was performed with the wav2vec 2.0 XLS-R  (0.3B) model~\cite{babu2021xls}. We followed the example in the Fairseq project toolkit~\cite{ott2019fairseq} to extract feature representations from self-supervised wav2vec 2.0 pre-trained model.\footnote{\url{https://github.com/pytorch/fairseq/tree/main/examples/wav2vec}}

\subsection{Fine-tuning}
Since pre-training is performed with only bona fide data (with no spoofed data), as per~\cite{wang2021investigating}, spoofing detection performance is expected to improve with fine tuning using in-domain bona fide and spoofed training data.  Our hypothesis is that fine tuning will protect against over-fitting and hence promote better generalisation.
For \emph{all} experiments reported in this paper, including those related to the ASVspoof 2021 LA dataset and the ASVspoof 2021 DF dataset, fine-tuning is performed using the ASVspoof 2019 LA training partition only. Whereas the 2021 LA data contains codec and transmission variation and the 2021 DF data contains compression variation, the 2019 LA data used for fine-tuning contains neither. During fine-tuning, the pre-trained wav2vec 2.0 XLS-R  model is optimsied jointly with the AASIST CM via back-propagation using the ASVspoof 2019 LA training set. This process is described in section~\ref{sec:Impl_details} whereas the fine-tuning procedure is illustrated to the right in Fig.~\ref{fig:wav2vec framework}. It is performed using a weighted cross entropy objective function to minimize the training loss. In contrast to pre-training, input masking is not applied to hidden features $z_{1:N}$ during fine-tuning. Additionally, we add a fully connected layer on top of the wav2vec 2.0 transformer encoder output $o_{1:N}$ in order to reduce the representation dimension (top-right of Fig.~\ref{fig:wav2vec framework}). 

\subsection{Use with AASIST CM}
The sinc layer shown in Fig.~\ref{fig:front-end framework}-(a) is replaced with the wav2vec 2.0 front-end shown in Fig.~\ref{fig:front-end framework}-(b).  
As before, the output $o_{1:N}$ is fed to a RawNet2-based residual encoder which is used to learn higher-level feature representations $\textbf{S}\in\mathbb{R}^{C \times F\times T}$. Whereas the baseline system extracts temporal and spectral representations $\textrm{\textbf{t}}$ and $\textrm{\textbf{f}}$ from $\textbf{\textrm{S}}$ using a max-pooling operation, a self-attentive aggregation layer (described in Section~5) was found to improve performance of both front-ends. As shown in Fig.~\ref{fig:baseline framework}, temporal and spectral representations are then fed to the AASIST CM model to obtain a two-class prediction (bona fide and spoofed) in the same manner as described in Section~\ref{sec:baseline}. A summary of the wav2vec 2.0 front-end and downstream AASIST model configurations is presented in Table~\ref{Tab:RawGAT details}.

\section{Self-attention based aggregation layer}
\label{sec:self_agg_layer}
Attention based pooling layers such as self-attentive pooling (SAP) and attentive statistical pooling (ASP)~\cite{okabe2018attentive} has shown to be beneficial to the aggregation of frame-level features and the extraction of embeddings~\cite{jung2020improving,jung2020graph,xie2019utterance,kim2021rawnext,shim2021graph} for speaker recognition and verification tasks.  We have also found that the introduction of a 2D self-attention based aggregation layer between the front-end and back-end helps to improve spoofing detection performance.

The new self-attentive aggregation layer is used to extract more attentive/relevant spectral and temporal representations. It helps to aggregate and assign higher attention weights through weighted summation to the most discriminative temporal and spectral features. We generate 2-D attention maps (an attention weight matrix) using a 2-D convolutional (conv2d) layer with \textit{one} kernel-size rather than conventional conv1d based attention applied to a single domain. Weights are derived from representations $\textbf{S}$ processed by a conv2d layer followed by an activation \& batch normalization (BN) layer, a 2-D convolutional layer and a softmax function to normalized the weights:
\begin{equation}
    W=\textrm{Softmax}(\textrm{conv2d}(\textrm{BN}(\textrm{SeLU}(\textrm{conv2d}(\textbf{S}))))),
    \label{Eq.: node propagation}
\end{equation}
\noindent where conv2d(·) denotes the 2-D convolution operation with an scaled exponential linear unit SeLU(·) as the activation function~\cite{klambauer2017self}, and BN is batch normalisation~\cite{Ioffe2015BatchShift}. Temporal and spectral representations are then extracted from the self-attentive aggregation layer according to:

\begin{equation}
\textrm{\textbf{t}}=\sum_{F} \textbf{S}\odot {W},
    \label{Eq:Temporal features}
\end{equation}
\begin{equation}
 \textrm{\textbf{f}}=\sum_{T} \textbf{S}\odot {W},
    \label{Eq:spectral features}
  \end{equation}
where $\odot$ denotes element-wise multiplication. $W \in\mathbb{R}^{F \times T}$ is the 2-D attention normalised learnable weight matrix used in the self-attentive aggregation layer to calculate the weighted \textit{sum} of the representation $\textbf{S}$ across time and frequency.

\begin{table}[!t]
\normalsize
	\centering
\renewcommand{\arraystretch}{1.3}
	\setlength\tabcolsep{1.4pt}
	\caption{The wav2vec 2.0 and AASIST model architecture and configuration.
	Dimensions refer to (channels, frequency, time).  Batch normalisation (BN) and scaled exponential linear unit (SeLU), beneath the dotted line, are applied to residual block outputs.}
	\vspace{0.2cm}
	%\resizebox{\linewidth}{!}
	%{%
	\begin{tabular}{p{2pt}p{2pt}p{2pt}p{2pt}}
	\hline
	\multicolumn{1}{c}{Layer} & \multicolumn{2}{c}{Input:64600 samples} & \multicolumn{1}{c}{Output shape}\\
	\hline\hline
		\multicolumn{1}{c}{Data-aug} & 
	\multicolumn{2}{c}{RawBoost} & \multicolumn{1}{c}{(64600)}\\
	\hline
	\multicolumn{1}{c}{SSL } & 
	\multicolumn{2}{c}{wav2vec 2.0 } & \multicolumn{1}{c}{(201,1024) (T,F)}\\
		\multicolumn{1}{c}{front-end} & 
	\multicolumn{2}{c}{ FC (fine-tuning)} & \multicolumn{1}{c}{(201,128)}\\
		\multicolumn{1}{c}{} & \multicolumn{2}{c}{transpose} & \multicolumn{1}{c}{$o$=(128,201) (F,T)}\\
\hline
%	\multicolumn{1}{c}{} & \multicolumn{2}{c}{FC(128)} & \multicolumn{1}{c}{(201,128)}\\

	\multicolumn{1}{c}{Post-} & \multicolumn{2}{c}{add channel} & \multicolumn{1}{c}{(1,128,201)}\\
	
	\multicolumn{1}{c}{processing} & \multicolumn{2}{c}{Maxpool-2D(3)} & \multicolumn{1}{c}{(1,42,67)}\\
	
	\multicolumn{1}{c}{} & \multicolumn{2}{c}{BN \& SeLU} & \multicolumn{1}{c}{}\\
	\hline
% 		\multicolumn{1}{c}{RawNet2-} & \multicolumn{2}{c}{Residual blocks} & \multicolumn{1}{c}{(64,42,67)}\\
		
% 		\multicolumn{1}{c}{encoder} & \multicolumn{2}{c}{BN \& SeLU} & \multicolumn{1}{c}{}\\
	\multicolumn{1}{c}{Res-block}2$\times$& 
       {${\left\{\begin{array}{c}
	   \text{Conv-2D((2,3),1,32)}\\
	    \text{BN \& SeLU }\\
	    \text{Conv-2D((2,3),1,32)}\\
	    \end{array}\right\} 
	    }$}&&\multicolumn{1}{c}{\hspace{2cm}}{(32,42,67)}\\
	\hline
	
	\multicolumn{1}{c}{Res-block}4$\times$&
        {${ \left\{ 
	    \begin{array}{c}
	   
	    \text{Conv-2D((2,3),1,64)}\\
	    \text{BN \& SeLU }\\
	    \text{Conv-2D((2,3),1,64)}\\
	    \end{array}\right\} 
	    }$}&&\multicolumn{1}{c}{\hspace{2cm}}{(64,42,67)}\\ 
	    \hdashline
	\multicolumn{1}{c}{} & \multicolumn{2}{c}{BN \& SeLU} & \multicolumn{1}{c}{}\\
	\hline
\multicolumn{2}{c}{Spectral-attention}& 
 \multicolumn{2}{c}{\hspace{3cm}}{Temporal-attention}\\ 
\hline
\multicolumn{2}{c}{self att. agg. layer (\textrm{\textbf{f}})=$(64,42)$} & 
\multicolumn{2}{c}{\hspace{2cm}}{self att. agg. layer (\textrm{\textbf{t}})=$(64,67)$}\\
\multicolumn{2}{c}{$\textrm{$\mathcal{G}_{s}$ = (21($N_\textrm{s}$), 64($d_s$))}$} &  \multicolumn{2}{c}{\hspace{2cm}}{$\textrm{$\mathcal{G}_{t}$ = (33($N_\textrm{t}$), 64($d_t$))}$}\\

% \multicolumn{2}{c}{$\textrm{graph module ($\mathcal{G}_{s}$) =(21,64)}$} &\hspace{1.55cm}  \multicolumn{2}{c}{$\textrm{graph module ($\mathcal{G}_{t}$) =(33,64)}$}\\
\hline 
 \multicolumn{1}{c}{hetero. graph $(\mathcal{G}_{st})$} & \multicolumn{2}{c}{HS-GAL } & \multicolumn{1}{c}{(54($N_\textrm{st}$),64($d_{st}$))}\\
 \hline 
\multicolumn{2}{c}{HS-GAL$\rightarrow$HS-GAL,}& 
 \multicolumn{2}{c}{\hspace{2cm}}{HS-GAL$\rightarrow$HS-GAL,}\\
% \multicolumn{2}{c}{stack node}&{\hspace{0.5cm}}\multicolumn{2}{c}{stack node}\\ 
\multicolumn{2}{c}{stack node}&\multicolumn{2}{c}{{\hspace{1.1cm}}stack node}\\ 
% \multicolumn{2}{c}{(32,26), (32,)}&{\hspace{2cm}}\multicolumn{2}{c}{(32,26),(32,)}\\ \cline{1-4}
\multicolumn{2}{c}{(32,26), (32,)}&\multicolumn{2}{c}{{\hspace{1.1cm}}(32,26),(32,)}\\ \cline{1-4}
\multicolumn{1}{c}{MGO $(\mathcal{G}_{ST})$} & \multicolumn{2}{c}{element-wise max.} & \multicolumn{1}{c}{(32,26), (32,)}\\
\hline
\multicolumn{1}{c}{readout} & \multicolumn{2}{c}{node-wise max. and avg.} & \multicolumn{1}{c}{(160,)}\\
\multicolumn{1}{c}{} & \multicolumn{2}{c}{\& concatenation} & \multicolumn{1}{c}{}\\
\hline	
\multicolumn{1}{c}{Output} & \multicolumn{2}{c}{FC(2)} & \multicolumn{1}{c}{2}\\	
\hline
\end{tabular}
%
%}
\label{Tab:RawGAT details}
\end{table}

\section{Experimental setup}

Described in the following are the databases and metrics used in all reported experimental work, our use of data augmentation and specific, reproducible implementation details.

\subsection{Databases and metrics}
\label{sec:data_aug}
We used the training and development partitions of the ASVspoof 2019 LA database~\cite{todisco2019asvspoof,wang2020asvspoof} for training and validation. Evaluation was performed using the ASVspoof 2021 LA and domain mis-matched DF databases~\cite{yamagishi2021_ASV_spoof}.  
While both are generated from the same VCTK source database\footnote{\url{http://dx.doi.org/10.7488/ds/1994}}, the LA database contains codec and transmission variability whereas the DF database contains compression variability in addition to data stemming from sources other than the VCTK database~\cite{vctk}. The 2019 data used for training and validation contains neither. We use two evaluation metrics: the Equal Error Rate (EER)~\cite{brummer2013bosaris} and the Minimum Tandem Detection Cost Function (min t-DCF)~\cite{kinnunen-tDCF-TASLP}.  We focus on the first for ease of interpretation and include the second since it is the default metric for ASVspoof challenges.

\subsection{Data augmentation}
\label{sec:DA}
Data augmentation (DA) is already known to 
reduce overfitting and hence to improve generalisation~\cite{das2021known,caceres2021biometric,tomilov2021stc,chen21b_asvspoof,tak2021rawboost} and is particularly effective in the case of LA scenarios in which there is substantial variability stemming from, e.g., encoding, transmission and acquisition devices~\cite{yamagishi2021_ASV_spoof}. 
We are interested to determine whether self-supervised learning is complementary to DA. Unlike traditional DA techniques which enlarge the training dataset using \emph{additional}, artificially generated utterances, and using the RawBoost\footnote{\url{https://github.com/TakHemlata/RawBoost-antispoofing}} DA tool~\cite{tak2021rawboost}, we add nuisance variability on-the-fly to the \emph{existing} training data. RawBoost adds variation in the form of: i)~linear and non-linear convolutive noise; ii)~impulsive signal-dependent additive noise; iii)~stationary signal-independent additive noise. Full details are available in~\cite{tak2021rawboost}.

DA is applied using exactly the same configuration and parameters reported in the original work~\cite{tak2021rawboost}.  It shows that a combination of linear and non-linear convolutive noise and impulsive signal-dependent additive noise strategies work best for the LA database.  These augmentation strategies suit the convolutive and device related noise sources that characterise telephony applications.  In contrast, for the DF database, DA works best using stationary signal-independent additive, randomly coloured noise, which match better the effects of compression~\cite{842996} applied in generating the DF database.  DA experiments and configuration are discussed further in Section~\ref{sec:DA_results}.

\subsection{Implementation details}
\label{sec:Impl_details}
Audio data are cropped or concatenated giving segments of approximately 4 seconds duration (64,600 samples). Graph pooling is applied with an empirically chosen pooling ratio of $k$ = 0.5 for spectral and temporal graphs. We used the standard Adam optimiser~\cite{kingma2014adam} with a fixed learning rate of 0.0001 for experiments without the wav2vec 2.0 front-end. Since fine-tuning demands high GPU computation, experiments with wav2vec 2.0 were performed with a smaller batch size of $14$ and a lower learning rate of $10^{-6}$ to avoid model over-fitting. As illustrated to right in Fig.~\ref{fig:wav2vec framework}, the fully connected layer after the wav2vec 2.0 SSL front end used 128 output dimensions. All other hyperparameters are the same for both front-ends which are both jointly optimised with the back-end classifier using back-propagation~\cite{Goodfellow_2016}. As is now common in the related literature~\cite{wang2021comparative,kanervisto2021optimizing}, we performed each experiments with three runs using different random seeds to initialize the network weights and report the results of the best performing seed and average results. All models were trained for 100 epochs on a single GeForce RTX 3090 GPU and all results are reproducible using open source code\footnote{URL for code to be made available upon publication.} and with the same random seed and GPU environment.

\section{Results}

We present five sets of experiments.  The first is a comparison of each front-end in terms of performance for the ASVspoof 2021 LA database.  The second and third assess the complementary benefits coming from the new self-attention based aggregation layer and data augmentation.  The fourth is an assessment performed on the ASVspoof 2021 DF database whereas the last is an assessment using a simplified CM solution.

\begin{table}[!t]
    \centering
     \caption{Pooled EER and pooled min~t-DCF results for the ASVspoof 2021 LA database evaluation set, for the sinc-layer and wav2vec 2.0 front-ends. SA refers to the self-attentive aggregation layer whereas DA refers to data augmentation.  Results are the best (average) obtained from three runs of each experiment with different random seeds.}
     \vspace{0.2cm}
    \label{tab:frontendLA}
      \renewcommand{\arraystretch}{1.1}
    \small
    \setlength\tabcolsep{3.75pt}
    \begin{tabular}{|c|c|c||c|c|}
      \hline
      front-end&SA&DA&Pooled EER&Pooled min t-DCF\\
      \hline\hline
     sinc-layer&$\times$&$\times$&11.47 (11.95)&0.5081 (0.5139)\\
     \hline
     wav2vec 2.0 &$\times$&$\times$&6.15 (6.46)&0.3577 (0.3587)\\
     \hline\hline
     sinc-layer&$\checkmark$&$\times$&8.73 (11.61)&0.4285 (0.5203)\\
     \hline
     wav2vec 2.0 &$\checkmark$&$\times$&4.48 (6.15)&0.3094 (0.3482)\\
 \hline\hline
    sinc-layer&$\checkmark$&$\checkmark$&7.65 (7.87)&0.3894 (0.3960)\\
     \hline
     \textbf{wav2vec 2.0 }&$\textbf{\checkmark}$&$\textbf{\checkmark}$&\textbf{0.82 (1.00)} & \textbf{0.2066 (0.2120)}\\
      \hline
    \end{tabular}
 
\end{table}

\subsection{Front-end comparison}
Results for the AASIST baseline with the sinc-layer front-end (Section~\ref{sec:baseline}) and the same system with the wav2vec 2.0 front-end (Section~\ref{sec:SSL_front_end}) are presented in the first two rows of Table~\ref{tab:frontendLA}.  These systems use neither the self-attentive aggregation layer nor data augmentation. The baseline EER of 11.47\% is high and shows that the system is not robust to the codec and transmission variability which characterises the ASVspoof 2021 LA dataset.  The same system using the wav2vec 2.0 front-end delivers an EER of 6.15\%.  While the relative reduction is almost 46\%, the EER is still unacceptably high.

\subsection{Self-attentive aggregation layer}

Results for the same two front-end variants but using the self-attentive aggregation layer (SA) introduced in Section~\ref{sec:self_agg_layer} are presented in rows 3 and 4 of Table~\ref{tab:frontendLA}.  In both cases the EER drops substantially, to 8.73\% for the sinc-layer front-end and to 4.48\% for the wav2vec 2.0 frontend.  In this case the wav2vec 2.0 frontend is responsible for a relative improvement of almost 50\%.

\subsection{Data augmentation}
\label{sec:DA_results}
Results for the same two systems, both with the self-attentive aggregation layer (SA), and now also with data augmentation (DA), are shown in rows 5 and 6 of Table~\ref{tab:frontendLA}. DA reduces the EER only marginally from 8.73\% to 7.65\% in case of the sinc-layer front-end. To verify that this improvement is not due to random factors in neural network training (e.g., different, random initial network weights), we conducted a statistical analysis of the results following~\cite{wang2021comparative}.
The results\footnote{Available in the form of an appendix.} 
suggest that the improvement is statistically significant and is hence unlikely to be caused by factors other than DA. Its effect is more pronounced when using the wav2vec 2.0 front-end for which the EER decreases from 4.48\% to 0.82\%.  This result is also statistically significant. This result corresponds to a relative improvement of almost 90\% when compared to the baseline EER of 7.65\%. To the best of our knowledge, this is the lowest EER reported for the ASVspoof 2021 LA database. 

\subsection{DeepFake results}

Results for exactly the same experiments, but for the ASVspoof 2021 DeepFake (DF) database, are shown in Table~\ref{tab:frontendDF}.    
While neither SA, nor DA improve upon the baseline EER of 21.06\%, consistency improvements are obtained for the wav2vec 2.0 front-end for which the EER drops from 7.69\% to 2.85\% using both SA and DA. This result is also statistically significant.  To the best of our knowledge, this is the lowest EER reported for the ASVspoof 2021 DF database. 
These results, while determined with the same wav2vec 2.0 front-end used for LA experiments, relate to a DA strategy optimised for the DF database (stationary signal-independent additive randomly coloured noise -- see Section~\ref{sec:DA}).  Results for \emph{exactly} the same setup, using the DA strategy optimised for LA (linear and non-linear convolutive noise and impulsive signal-dependent additive noise) are shown in the last two rows of Table~\ref{tab:frontendDF}.  While the EER increases to 6.64\%, this is still a competitive result and is 67\% lower relative to the result of 20.04\% for the sinc-layer front-end. Whereas a component of the DF database originates from the same VCTK database as the entire LA database, other components are sourced from multiple different corpora (Voice Conversion Challenge 2018 and 2020 databases)~\cite{yamagishi2021_ASV_spoof} including spoofed utterances generated with more than 100 different algorithms.  With the ASVspoof 2019 LA training data containing neither codec or transmission variability (LA evaluation data), nor compression variability (DF evaluation), results show that the use of better pre-trained models leads to consistent improvements in generalisation, here being previously unseen spoofing attacks.  Results for the DF database show that the benefit extends also to the case of domain mismatch.

\begin{table}[!t]
    \centering
     \caption{As for Table~\ref{tab:frontendLA} except for the ASVspoof DF database, evaluation set.  Since there is no ASV in the DF scenario, there are no min t-DCF results.  The last two lines show results for an LA-optimised DA configuration. }
     \vspace{0.2cm}
  
    \label{tab:frontendDF}
      \renewcommand{\arraystretch}{1.1}
    \small
    \setlength\tabcolsep{13pt}
    \begin{tabular}{|c|c|c||c|}
      \hline
      front-end&SA&DA&Pooled EER\\
      \hline\hline
      sinc-layer&$\times$&$\times$&21.06 (22.11)\\
     \hline
     wav2vec 2.0 &$\times$&$\times$&7.69 (9.48) \\
     \hline\hline
     sinc-layer&$\checkmark$&$\times$&23.22 (25.08)\\
     \hline
     wav2vec 2.0 &$\checkmark$&$\times$&4.57 (7.70)\\
     \hline\hline
    
     sinc-layer&$\checkmark$&$\checkmark$&24.42 (25.38)\\
     \hline
    \textbf{ wav2vec 2.0} &$\textbf{\checkmark}$&$\textbf{\checkmark}$& \textbf{2.85 (3.69)}\\
     \hline\hline
     sinc-layer&$\checkmark$&$\checkmark\renewcommand{\thefootnote}{\fnsymbol{footnote}}\footnote[1]{\label{fn:DA}Results using the DA strategy optimised for LA (linear and non-linear convolutive noise and impulsive signal-dependent additive noise).}$&20.04 (20.50)\\
     \hline
     wav2vec 2.{0}&$\checkmark$&$\checkmark\textsuperscript{\ref{fn:DA}}$& 6.64 (7.32)\\
     
    \hline
     
  \end{tabular}

\end{table}
\subsection{Simplified CM solution}
The last set of experiments were performed in order to gauge the relative importance of the AASIST and whether the improvements in generalisation are obtained for a simpler CM solution. We removed the RawNet2-based encoder and replaced AASIST with a simple back-end comprising a max-pooling layer, a single graph layer and a single linear layer. Results for both ASVspoof 2021 LA and DF databases using optimised DA strategies for each are shown in Table~\ref{tab:SSL_simple_back_end}. LA and DF results of 1.19\% and 4.38\% show that competitive EERs can obtained using the fine-tuned wav2vec 2.0 front-end even with relatively less complex networks and that the benefits to generalisation are still complementary to those of DA.
\begin{table}[!t]
    \centering
     \caption{Pooled EER and pooled min~t-DCF (LA only) results for the ASVspoof 2021 LA and DF databases (DBs), evaluation sets, using DB-optimised DA and the simplified back-end. }
     \vspace{0.2cm}
    \label{tab:SSL_simple_back_end}
      \renewcommand{\arraystretch}{1.1}
    \small
    \setlength\tabcolsep{4pt}
    \begin{tabular}{|c|c|c||c|c|}
      \hline
      front-end&DA&DB&Pooled EER&Pooled min t-DCF\\
      \hline\hline
      
       wav2vec 2.0 &$\checkmark\renewcommand{\thefootnote}{\fnsymbol{footnote}}\footnote[2]{\label{fn:DA_algo}We used their respective optimised DA strategies for LA and DF as described in Section~\ref{sec:DA}}$&LA&1.19&0.2175\\
      \hline

      wav2vec 2.0 &$\checkmark\textsuperscript{\ref{fn:DA_algo}}$&DF&4.38 &-\\
    \hline
     
  \end{tabular}

\end{table}

\section{Conclusions and discussion}
 We report in this paper our attempts to harness the power of self-supervised learning in the form of the popular wav2vec 2.0 front-end to learn more reliable representations to improve spoofing detection performance. We show that a well-trained, fine-tuned front-end, even when learned initially using massive quantities of only \emph{bona fide} utterances, can improve generalisation substantially.  Compared to a sinc-layer front-end, when coupled with a new self-attentive aggregation layer and data augmentation, the wav2vec 2.0 front-end delivers up to a 90\% relative reduction in the equal error rate for the logical access spoofing detection task and up to an 88\% relative reduction for a domain mis-matched Deepfake detection task in which
spoofed utterances are generated with more than 100 different attack algorithms. Improvements stemming from the use of the self-supervised wav2vec 2.0 front-end are consistent for every experimental comparison and, to the best of the authors' knowledge, results are the lowest reported EERs for both LA and DF databases to date.

We must nonetheless acknowledge that almost all results reported in the literature are derived using \emph{fixed} training data, whereas those reported in this paper are derived from a model pre-trained using additional \emph{external} data.  Given that our results are obtained with different systems trained using different data, comparisons to the majority of results reported in the literature are obviously not fair.  The scale of the improvements, however, up to a 90\% relative reduction in EER, indicate the potential gain in performance that can be obtained with the use of additional, external training data and might suggest that the training data restrictions for ASVspoof evaluations might be relaxed.  Nonetheless, the wav2vec 2.0 model is massively more complex when compared to the previous state-of-the-art solutions. Whether or not  solutions with such footprints can be adapted to support practical applications remains to seen. 

Future work should investigate which particular characteristics of the self-supervised front-end are most beneficial.  One objective of such work is to use the results of such analysis as a starting point to scale down the model size and complexity so that it might be more easily adopted for practical scenarios with modest computational resources.  

\vfill

\bibliographystyle{IEEEbib}
\bibliography{Odyssey2022_BibEntries}

\begin{appendices}
\begin{figure*}
 \vspace{-2.05cm}
 \section{Statistical analysis results}
  \centering
  \includegraphics[trim={2.4cm 2cm 2cm 0cm},clip,width=1.175\linewidth]{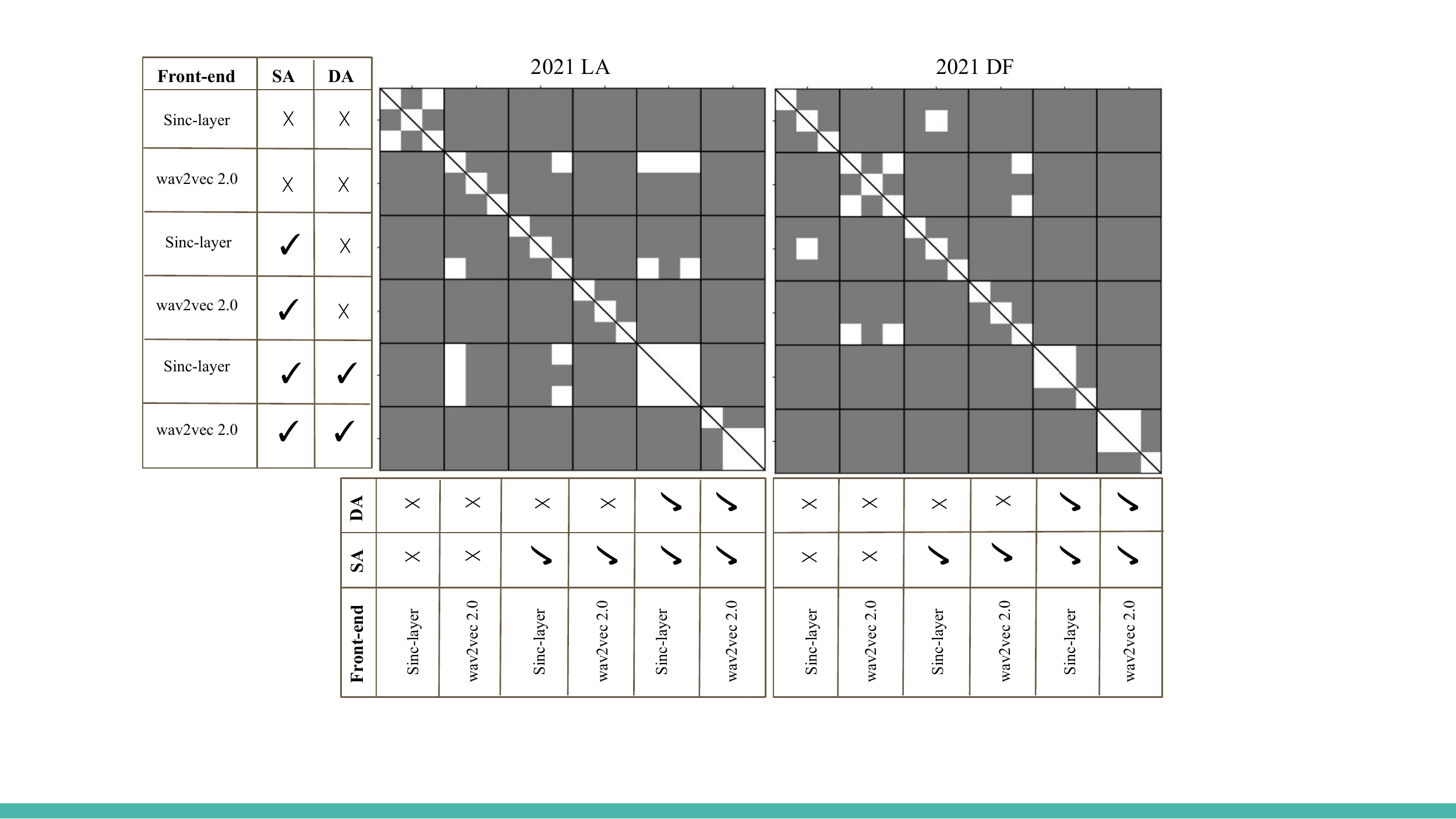}
 \caption{Statistical significance test using EERs on LA and DF 2021 evaluation dataset and Holm-Bonferroni correction with $\alpha$ = 0.05. Gray colour indicates significant difference, and insignificant difference indicates by white colour. Each square in the black square contains 3 × 3 entries and denotes pair-wise tests between three training-evaluation runs of two models. Front-end systems are in the same order as that in Table 2 and Table 3.}
 \end{figure*}
 
 \begin{table*}[!t]
 \vspace{-3cm}
  \section{Breakdown EER's (\%) pooled over attacks on ASVspoof 2021 LA and DF databases}
    \centering
     \caption{Breakdown results in terms of EERs (\%) for each codec conditions (C1-C7) of Table 2 on ASVspoof 2021 LA evaluation partition. Results are the best (average) obtained from three runs of each ex-periment with different random seeds.}
      \renewcommand{\arraystretch}{2}
    \small
    \setlength\tabcolsep{7pt}
   \begin{tabular}{|c|c|c||c|c|c|c|c|c|c||c|c|}
      \hline
      front-end&SA&DA&C1&C2&C3&C4&C5&C6&C7& Pooled EER & Pooled min t-DCF 	\\ 
     \hline \hline
     Sinc layer&$\times$&$\times$&6.36  &  6.76  &  15.64  &   7.85  &  6.37   & 10.40  &  12.39   & 11.47 (11.95) & 0.5081 (0.5139)\\
     \hline 
     wav2vec 2.0&$\times$&$\times$&2.06  &   3.31 &   19.09  &  2.79  &   3.18   &  7.55  &  4.63 &6.15 (6.46)&0.3577 (0.3587)\\
     \hline
     
      Sinc layer&$\checkmark$&$\times$&2.36 &   5.25  &  16.72 &   3.09 &   4.80 &    6.66 &    4.97&8.73(11.61)&0.4285 (0.5203)\\
     \hline 
 
    wav2vec 2.0&$\checkmark$&$\times$&0.26 &   1.19  &   7.84  &  0.77 &   1.02   &  4.97 &   2.89 &4.48 (6.15)&0.3094 (0.3482)\\
     \hline
     Sinc layer&$\checkmark$&$\checkmark$& 3.75  &   5.77  &   6.87  &   4.61  &  5.57   &  6.58   &  7.08 &7.65 (7.87)&0.3894 (0.3960)\\
    \hline 
     wav2vec 2.0&$\checkmark$&$\checkmark$&0.30  &  0.58   & 0.56  &  0.67 &   0.52 &   0.81  &  0.98&0.82 (1.00)&0.2066 (0.2120)\\
     \hline 
  \end{tabular}
   \label{tab:breakdown LA results}
\end{table*}
\begin{table*}[!b]
    \centering
    \vspace{-8cm}
     \caption{Breakdown results in terms of EERs (\%) for each codec conditions (DFC1-DFC8) of Table 3 on ASVspoof 2021 DF evaluation partition.}
      \renewcommand{\arraystretch}{2}
    \setlength\tabcolsep{7pt}
    \begin{tabular}{|c|c|c||c|c|c|c|c|c|c|c||c|}
      \hline
      front-end&SA&DA&DF-C1&DF-C2&DF-C3&DF-C4&DF-C5&DF-C6&DF-C7&DF-C8& Pooled EER  \\ 
     \hline \hline
     Sinc layer&$\times$&$\times$&23.53  &   24.74   &   24.05 &      23.84   &   19.06   &   17.58   &   18.49  &   19.06 &21.06 (22.11)\\
      \hline 
     wav2vec 2.0&$\times$&$\times$&8.21  &     9.04   &    8.02 &       8.11    &   6.15    &   6.31     &  6.93    &  6.06 &7.69 (9.48)\\
     \hline
    
     Sinc layer&$\checkmark$&$\times$&26.98   &   28.25    &  27.84   &    27.40     & 19.93   &   19.32   &   20.11 &    19.66  &23.22 (25.08) \\
     \hline 
     wav2vec 2.0&$\checkmark$&$\times$& 4.58 &     4.84   &    4.07    &    4.28   &    3.41   &    3.71   &    4.10  &    3.53  &4.57 (7.70)\\
     \hline
    
     Sinc layer&$\checkmark$&$\checkmark$& 28.81   &   29.12   &   29.07       &28.38    &  21.11    &  20.50   &   21.02     &20.89&24.42 (25.38)\\
    \hline 
     wav2vec 2.0&$\checkmark$&$\checkmark$&2.34    &   4.30     &  2.64        &2.37   &   2.58     &   1.92      & 3.31    &  2.75 &2.85 (3.69)\\
     \hline\hline
  
    Sinc layer&$\checkmark$&$\checkmark\renewcommand{\thefootnote}{\fnsymbol{footnote}}\footnote[1]{\label{fn:DA}Results using the DA strategy optimised for LA (linear and non-linear convolutive noise and impulsive signal-dependent additive noise).}$&23.98 &     24.56   &   24.83   &    23.67   &   17.01  &    16.39    &  17.23 &    16.48 &20.04 (20.50)\\
    \hline
    wav2vec 2.0 &$\checkmark$&$\checkmark\textsuperscript{\ref{fn:DA}}$  & 7.88&5.62&7.05&7.51&5.18&6.06&5.22&5.36&6.64 (7.32)\\
     \hline

  \end{tabular}
   
    \label{tab:breakdown DF results}
    \textcolor{red}{$^*$}Results using the DA strategy optimised for LA (linear and non-linear convolutive noise and impulsive signal-dependent additive noise).
\end{table*}

 \end{appendices}

\end{document}